\documentclass{ws-ijmpd}
\usepackage[super,compress]{cite}
\usepackage{amssymb,amsmath,amscd}
\usepackage{graphicx,calc,epsfig,pstricks,bbm}
\usepackage{tikz}

\newcommand {\be}{\begin{equation}}
\newcommand {\ee}{\end{equation}}
\newcommand{\ba}{\begin{array}{c}}
\newcommand{\ea}{\end{array}}

\newcommand{\ket}[1]{|#1\rangle}

\newcommand{\scr}{\scriptscriptstyle}
\newcommand{\vgraph}{\mathfrak{n}}
\newcommand{\cube}{\ba
 \begin{tikzpicture}
\pgfmathsetmacro{\cubex}{0.15}
\pgfmathsetmacro{\cubey}{0.15}
\pgfmathsetmacro{\cubez}{0.15}
\draw (0,0,0) -- ++(-\cubex,0,0) -- ++(0,-\cubey,0) -- ++(\cubex,0,0) -- cycle;
\draw (0,0,0) -- ++(0,0,-\cubez) -- ++(0,-\cubey,0) -- ++(0,0,\cubez) -- cycle;
\draw (0,0,0) -- ++(-\cubex,0,0) -- ++(0,0,-\cubez) -- ++(\cubex,0,0) -- cycle;
\end{tikzpicture}
\ea}

\begin{document}

\markboth{E. Alesci, F. Cianfrani}
{Quantum Reduced Loop Gravity...}

%
\catchline{}{}{}{}{}
%

\title{QUANTUM REDUCED LOOP GRAVITY AND THE FOUNDATION OF LOOP QUANTUM COSMOLOGY}

\author{EMANUELE ALESCI}

\address{Instytut Fizyki Teoretycznej, Uniwersytet Warszawski, ul. Pasteura 5, 02-093 Warszawa, Poland.\\
SISSA, Via Bonomea 265, 34126 Trieste, Italy and INFN Sez. Trieste.\\
emanuele.alesci@fuw.edu.pl}

\author{FRANCESCO CIANFRANI}

\address{Institute
for Theoretical Physics, University of Wroc\l{}aw, Plac\ Maksa Borna
9, Pl--50-204 Wroc\l{}aw, Poland.\\
francesco.cianfrani@ift.uni.wroc.pl}

\maketitle

\begin{history}
\received{Day Month Year}
\revised{Day Month Year}
\end{history}

\begin{abstract}
Quantum Reduced Loop Gravity is a promising framework for linking Loop Quantum Gravity and the effective semiclassical dynamics of Loop Quantum Cosmology. We review its basic achievements and its main perspectives, outlining how it provides a quantum description of the Universe in terms of a cuboidal graph which constitutes the proper framework for applying loop techniques in a cosmological setting. 
\end{abstract}


\ccode{PACS number: 04.60.Pp}


\section{Introduction}	

Cosmology is probably the hottest topic for a quantum theory theory of gravity. Cosmological observations are becoming more and more 
accurate and we have entered into the era of precision cosmology. Hence, the best perspectives for eventual experimental confirmations 
of a quantum gravity scenario are going to be realized in cosmology, with the forthcoming missions on cosmic microwave background spectrum 
analysis, the advent of neutrino cosmology and maybe also with the detection of gravitational waves. Furthermore, the unsolved issues in cosmology 
(big bang, inflation, baryogenesis, dark matter, dark energy, ..) call for a theoretical effort to explain them from first principles and 
quantum gravity may provide new insights. However, the quantum gravity problem is very complicated both on a technical point of view and from an interpretative perspective. In cosmology, while the interpretative issues stay the same (or are even worse), the technical analysis is much easier and some quantization 
techniques can be successfully applied. In this respect, it is worth noting the case of Loop Quantum Cosmology (LQC) \cite{Bojowald:2011zzb,Ashtekar:2011ni} , which performs the quantization of the minisuperspace models describing homogeneous spaces and infers a singularity free description of the Universe history. The phenomenology of LQG is nowadays a well-established subject of investigation and several interesting achievements has been obtained: from the replacement of the big bang with a big bounce \cite{Singh:2006im,Ashtekar:2006rx,Ashtekar:2006wn,Ashtekar:2007em} , to the predicted modification to the CMB spectrum \cite{Bojowald:2011hd,Barrau:2014maa,Ashtekar:2015dja} and the set up of initial conditions for inflation \cite{Singh:2006im,Corichi:2010zp,Ashtekar:2011rm,Agullo:2012sh,Agullo:2013ai} (see \cite{Gupt:2013swa} for the extension to Bianchi type I). 

In view of these promising phenomenological results, the investigation on the foundation of LQC is crucial for both the internal consistency of the model and the kind of information one can extract from cosmology for the full theory, namely Loop Quantum Gravity (LQG) \cite{Rovelli:2004tv,Thiemann:2007zz,Ashtekar:2004eh} . The realization of a direct link between LQG and LQC is the aim of Quantum Reduced Loop Gravity (QRLG) (see \cite{Brunnemann:2005in,Brunnemann:2005ip}
for early attempts towards the same goal and \cite{Gielen:2013kla,Gielen:2014uga,Calcagni:2014tga} for a similar idea in the framework of Group Field Theory). 

LQC is a minisuperspace quantization, in which homogeneity and, eventually, isotropy are classically implemented and the resulting dynamic system is quantized in polymer representation. The Hamiltonian is ambiguous, due to the arbitrariness of the polymer parameter which acts as a regulator. This has been originally fixed as a constant, the so-called $\mu_0$ scheme, but it was then realized how some issues are present and a different choice, the $\bar\mu$ scheme, is preferable. 
The main ingredients of LQG seem to be lost in LQC: the graph and $SU(2)$ gauge structures. Indeed, the former can be recovered ``a posteriori'' in lattice-refined LQC \cite{Bojowald:2006qu}, in which a collection of $N$ LQC-universes is considered, showing how the resulting collective dynamics is characterized by a polymer parameter proportional to $N^{-1/3}$. 

In QRLG the kinematics is defined before the reduction to minisuperspace, such that the main features of LQG (graph structure and SU(2) quantum numbers) are present, but in a simplified context such that the relevant computations can be performed analytically. The main ideas is to implement on a quantum level (namely in the kinematical Hilbert space of LQG) the gauge-fixing conditions restricting to diagonal spatial metric and triads \cite{Alesci:2012md,Cianfrani:2013pwa,Alesci:2013lea,Alesci:2013xya}. The minisuperspace reduction is performed at the dynamic level, keeping only those terms preserving the diagonal metric condition. Such a reduction provides a suitable framework for the investigation of the Bianchi I model, whose scalar constraint can be defined as an analytic operator in the Hilbert space of QRLG \cite{Alesci:2013xd}. The investigation on the Bianchi I case leads to the construction of a proper semiclassical limit \cite{Alesci:2014uha}, which outlines how there is a correspondence between the effective semiclassical description of QRLG and the quantum dynamics of LQC. The analysis of collective modes \cite{Alesci:2014rra,Alesci:2015jca,Alesci:2015nja} (Bianchi I patches with several nodes) fixes definitively such a correspondence and determines the origin of the regulator according with the prediction of lattice refined LQC. However, the number of nodes is fixed in QRLG, thus the theory reproduces the $\mu_0$ scheme.

In what follows, we present an introduction to the formalism of QRLG. At first, we give a very brief introduction to LQG and to its main tools. Then, we analyze the structure of the reduced phase space for the Bianchi I model and we quantize the resulting system according with LQC, pointing out how it can be given an effective semiclassical description capturing the relevant modification to the classical dynamics. Hence, we define the main tools of QRLG: the implementation of the gauge fixing conditions and the truncation of the dynamics. We outline how the scalar constraint operator can be defined and its semiclassical limit reproduces the effective semiclassical Hamiltonian one can introduce for LQC. Then, this formulation is extended to the case in which a scalar field is present. 
Finally, the remaining issues and the main perspectives are discussed. 

\section{Loop Quantum Gravity in a nutshell}

The canonical quantization program for gravity substantially advanced thanks to the contribution of LQG. In fact, the basic result of such quantum gravity approach is the definition of a Hilbert space structure on which the constraints of the theory can be defined and eventually solved and the states represent discrete (dual) geometries. This is obtained by using a different parametrization of the phase space with respect to the metric formulation and by adopting background-independent quantization tools. 
In particular, the phase space of gravity can be parametrized by Ashtekar-Barbero connections $A^i_a$ and their conjugate variables $E^i_a$, which are inverse densitized triads of the spatial metric. The merit of this formulation is that the Gauss constraint of a $SU(2)$ gauge theory comes out, $A^i_a$ being the associated connection, and some technicalities originally developed for Yang-Mills theories can be adopted, as for instance the use of holonomies along graphs instead of connections at points. In fact, the phase space of LQG is described by the holonomies of Ashtekar-Barbero connections $A^i_a$ and the fluxes of $E^i_a$ across surfaces, such that Poisson brackets are finite. The holonomy-flux algebra can be quantized and a representation can be given on cylindrical functions. The resulting kinematical Hilbert space $\mathcal{H}$ is the direct sum over all piece-wise graph $\Gamma$ of that based at a single graph $\Gamma$, $\mathcal{H}_\Gamma$, whose elements are functions of $L$ copies of the $SU(2)$ group, $L$ being the total number of links in $\Gamma$. Since each $SU(2)$ function can be expanded in irreducible representations of the $SU(2)$ group, basis vectors are simply given by
 \be
<h |\Gamma,\{j_l\}>= \prod_{l}D^{j_{l}}(h_{l}),
\label{spinnet solite0} 
\ee
in which at each link $l$ the Wigner matrices $D^{j_{l}}(h_{l})$ in the representation $j_{l}$ has been inserted and the product extends over all the links $l$ in $\Gamma$. 

Holonomy operators acts according with the composition rule for the $SU(2)$ group. Hence, they need to be expanded in irreps and the action of each irreps can be inferred from $SU(2)$ recoupling theory

\be
\begin{array}{c}
\ifx\JPicScale\undefined\def\JPicScale{1}\fi
\psset{unit=\JPicScale mm}
\psset{linewidth=0.3,dotsep=1,hatchwidth=0.3,hatchsep=1.5,shadowsize=1,dimen=middle}
\psset{dotsize=0.7 2.5,dotscale=1 1,fillcolor=black}
\psset{arrowsize=1 2,arrowlength=1,arrowinset=0.25,tbarsize=0.7 5,bracketlength=0.15,rbracketlength=0.15}
\begin{pspicture}(0,0)(25,20)
\psline(12,9)(12,1)
\psline{-<<}(12,5)(5,5)
\psline{<-}(25,5)(18,5)
\psline(12,9)(18,5)
\psline(18,5)(12,1)
\psline(12,20)(12,12)
\psline{-<<}(12,16)(5,16)
\psline{<-}(25,16)(18,16)
\psline(12,20)(18,16)
\psline(18,16)(12,12)
\rput(22,3){$j_1$}
\rput(22,14){$j_2$}
\end{pspicture}
\end{array}
=
\sum_k 
\begin{array}{c}
\ifx\JPicScale\undefined\def\JPicScale{1}\fi
\psset{unit=\JPicScale mm}
\psset{linewidth=0.3,dotsep=1,hatchwidth=0.3,hatchsep=1.5,shadowsize=1,dimen=middle}
\psset{dotsize=0.7 2.5,dotscale=1 1,fillcolor=black}
\psset{arrowsize=1 2,arrowlength=1,arrowinset=0.25,tbarsize=0.7 5,bracketlength=0.15,rbracketlength=0.15}
\begin{pspicture}(0,0)(26,10)
\psline(12,9)(12,1)
\psline{-<<}(9,5)(2,2)
\psline{<-}(26,8)(21,5)
\psline(12,9)(18,5)
\psline(18,5)(12,1)
\psline{-<<}(9,5)(2,8)
\psline{<-}(26,2)(21,5)
\rput(3,0){$j_1$}
\rput(2,10){$j_2$}
\psline(9,5)(12,5)
\psline(18,5)(21,5)
\rput(24,9){$j_2$}
\rput(24,1){$j_1$}
\rput(10,3){$k$}
\end{pspicture}
\end{array}\label{su2rec}
\ee

where we introduced a useful graphical notation, in which triangles denote $SU(2)$ group elements, the labels $j_1,j_2$ refer to the spin number of the considered representation along the line and the three-valent nodes represent Clebsch-Gordan coefficients. The sum extends over all the admissible representations $k$, {\it i.e.} $|j_1-j_2|\leq j\leq j_1+j_2$. 

The action of fluxes $E_i(S)$ across a surface $S$ is obtained from the requirement that they realize a faithful representation of the holonomy-flux algebra. They act as left (right)-invariant vector fields of the SU(2) group: given a surface $S$ having a single intersection with $\Gamma$ in a point $P\in l$, such that $l=l_1\bigcup l_2$ and $l_1\cap l_2=P$, the operator $\hat{E}_i(S)$ provides the insertion of the $SU(2)$ generator $\tau_{i}$ in $P$, {\it i.e.}
\be
\hat{E}_i(S)D^{(j_l)}(h_l)
=8\pi\gamma l_P^2 \; o(l,S) \; D^{j_l}(h_{l_1})\,\tau_{i}\,D^{j_l}(h_{l_2}),\label{Eop}
\ee  
$\gamma$ and $l_P$ denoting the Immirzi parameter and the Planck length, respectively, while $o(l,S)$ is equal to $0,1,-1$ according with the relative sign of $l$ and the normal to $S$.  

At this point, according with Dirac prescription we need to implement on a quantum level the constraints, which are the $SU(2)$ Gauss constraint, the vector and scalar constraint. Let us start with the $SU(2)$ Gauss constraint: it generates $SU(2)$ transformations, which act at the beginning and ending points of each link. Hence, a gauge-invariant state is obtained by inserting at nodes $\vgraph$ invariant intertwiners $x_\vgraph$, mapping the sum of the representations at the links emanating from $\vgraph$ into the trivial (gauge-invariant) representation. Therefore, basis elements of the $SU(2)$-invariant Hilbert space are
\be
<h |\Gamma,\{j_l\},\{x_\vgraph\}>=\prod_{\vgraph\in\Gamma} {x_{n}}\cdot   \prod_{l}D^{j_{l}}(h_{l}),
\label{spinnet solite} 
\ee
where the products extend over all the nodes $\vgraph$ in $\Gamma$ and all the links $l$ emanating from $\vgraph$. The symbol $\cdot$ means the contraction between the magnetic indexes of the intertwiners and of Wigner matrices.

The vector constraint cannot be defined in the kinematical Hilbert space, the reason being that the infinitesimal variation of a holonomy under diffeomorphisms is not an holonomy anymore. However, we can define diffeo-invariant states, which are invariant under finite diffeomorphisms, by introducing s-knots \cite{Ashtekar:1995zh} , {\it i.e.} equivalence class of spin networks under diffeomorphisms. A rigorous definition of these states can be given in the dual space $\mathcal{H}^*$. 

The last constraint is the hardest, the scalar constraint. It can be regularized \cite{Thiemann:1996aw} via a graph-dependent triangulation $T$. Here the graph is that of the states on which the operator acts and the triangulation contains all the tetrahedra $\Delta$ obtained by considering all the triples of links emanating from any given node $n$. The resulting expression can be split into two parts: the Euclidean and Lorentian part. For instance, the expression for the Euclidean part of the scalar constraint reads 
 \be
   \label{Hm_delta:quantum}
 \hat{\mathcal{S}}_E=\sum_{\Delta\in T} \hat{H}^m_{\Delta}[N]:= \sum_{\Delta\in T}\, \mathcal{N}(n)C(m) \,  \, \epsilon^{ijk} \,
   \mathrm{Tr}\Big[\hat{h}^{(m)}_{\alpha_{ij}} \hat{h}^{(m)-1}_{s_{k}} \big[\hat{h}^{(m)}_{s_{k}},\hat{V}\big]\Big] .
 \ee
$V$ being the volume operator, acting at nodes and depending on the intertwiner, while the index $m$ refers to the adopted representation of the holonomies entering the expression above (in what follows we will consider $m=1/2$). $C(m)=\frac{-i}{8\pi \gamma l^2_P (d_m\, m (m+1))^2_m}$ denotes a normalization constant and $\mathcal{N}$ is the lapse function. $\alpha_{ij}$ is the perimeter of the base of the tetrahedron $\Delta$ and it is made of two segments $s_i$ and $s_j$ belonging to two links $l_i$ and $l_j$ emanating from $n$, while $s_k$ is the segment belonging to the remaining element of the triple of links, $l_k$.

The volume operator is a very complicated object, since it is the square root of a modulus, thus a double square root of a square. No analytic expression for such an object has been obtained, such that no analytic expression for the scalar constraint operator exists and only some formal solutions in terms of ``dressed nodes'' \cite{Thiemann:2007zz,Thiemann:1996aw} have been found. 

It is worth noting how a different graph-dependent triangulation can be adopted (as those presented in \cite{Alesci:2010gb}), for instance by replacing tetrahedra with cubes, the only difference being the shape of $\alpha_{ij}$ in \eqref{Hm_delta:quantum}. This is what we will do for QRLG.

\section{Reduced phase space and Loop Quantum Cosmology}

The Bianchi I model is the simplest anisotropic cosmological space and its line-elements reads
\be
ds^2=\mathcal{N}^2(t)\, dt^2-dl^2\qquad dl^2=(a_1(t))^2\, (dx_1)^2+(a_2(t))^2\, (dx_2)^2+(a_3(t))^2\, (dx_3)^2\,,\label{B1}
\ee
where $a_i$ $i=1,2,3$ denote the scale factors. The space is homogeneous and metric components do not depend on spatial coordinates $x^i$. This is not the case anymore if we perform a generic coordinate transformation. Hence, the restriction to homogeneous scale factors and lapse function entails the choice of a privileged reference frame, thus the breakdown of manifest diffeomorphisms invariance (it is only a formal breakdown, since a posteriori it can be verified that the supermomentum or the vector constraints identically vanish).

Reduced phase space variables in metric formulation are the three scale factors with the associated conjugate momenta, while the lapse function is a Lagrange multiplier. The corresponding variables in LQG are obtained by evaluating $A^i_a$ and $E^a_i$ for the spatial metric in \eqref{B1}. This is usually done by fixing spatial triads as follows 
\be
e^i_a= a_i\,\delta^i_a\,,\label{tetr}
\ee
where the repeated index $i$ is not summed. This choice is arbitrary, since any triad related to \eqref{tetr} by a rotation on internal indexes is equally admissible. Since such internal rotations are generated by the $SU(2)$ Gauss constraint, we performed a gauge-fixing of the associated symmetry \cite{Cianfrani:2010ji,Cianfrani:2012gv}. Therefore, in reduced phase space we have already lost the two kinematical symmetries of the theory: background independence and $SU(2)$ gauge invariance. 

Once the choice \eqref{tetr} is done and a fiducial volume $\mathcal{V}_0=\ell_0^3$ is considered, $A^i_a$ and $E^a_i$ are given by
\be
A^i_a=(\ell_0)^{-1}\, c_i\,\delta^i_a\qquad E^a_i=(\ell_0)^{-2}\, p_i\,\delta^a_i\,,
\ee
where $c_i$ and $p_i$ are the new coordinates in reduced phase space and their explicit expression in terms of the scale factors are
\be
c_i= \ell_0\, \frac{\gamma}{N}\,\dot{a}_i \qquad p_i=(\ell_0)^2 \, \frac{a_1 a_2 a_3}{a_i}\,,\label{pc}
\ee
$\dot{\phantom1}$ denoting time derivative. The Poisson brackets are canonical up to a constant factor 
\be
\{c_i, p_i\}=\frac{8\pi G\gamma}{3}\,.
\ee
The only constraint is the scalar one $\mathcal{S}$: the Lorentzian and Euclidean parts are proportional and provide the following expression
\begin{equation}
\mathcal{S}=\frac{1}{\gamma^2\mathcal{V}_0}\left(\sqrt{\frac{p_1p_2}{p_3}}\,c_1c_2+\sqrt{\frac{p_2p_3}{p_1}}\,c_2c_3+\sqrt{\frac{p_3p_1}{p_2}}\,c_3c_1\right)=0\,,\label{S} 
\end{equation}
while the full Hamiltonian is just proportional to it:
\begin{equation}
H=\frac{N\mathcal{V}_0}{8\pi G}\mathcal{S}\,.\label{Ham}
\end{equation}

In LQC \cite{MartinBenito:2008wx,Ashtekar:2009vc} , the quantization is performed starting from the analogous of the holonomy-flux algebra for the variables $\{c_i,p_i\}$, which are $\{N_{\vec{\mu}}, p_i\}$
with $N_{\vec\mu}$ the product of three quasi-periodic functions of $c_i$
\be
N_{\vec\mu}=e^{i\mu_1 c_1}\,e^{i\mu_2 c_2}\,e^{i\mu_3 c_3}\,,
\ee
$\vec{\mu}=(\mu_1,\mu_2,\mu_3)$ being a triple of real numbers. These are the wave functions for the resulting quantum states $|\vec{\mu}\rangle$ and the operators act as follows
\be
\hat{N}_{\vec{\nu}}|\vec{\mu}\rangle=|\vec{\nu}+\vec{\mu}\rangle\qquad \hat{p}_i|\vec{\mu}\rangle= \frac{8\pi\gamma\ell_P^2}{3}\,\mu_i|\vec{\mu}\rangle\,,
\ee
while the scalar product reads
\be
\langle\vec{\nu}|\vec{\mu}\rangle=\delta_{\vec{\nu},\vec{\mu}}\,.
\ee
The full Hilbert space is the direct product of three Bohr compactifications of the real line
\be
\mathcal{H}^{LQC}=\otimes_{i=1}^3 \mathcal{L}^2(\mathbb{R}^i_{Bohr},d\mu^i_{Bohr})\,.
\ee
As for connections in LQG, the operators associated to $c_i$ do not exist. Hence, the operator associated to $\mathcal{S}$ is defined by fixing minimum values $\bar\mu$'s for $\mu$'s, such that one can make the replacements 
\begin{align}
&c_i\rightarrow \frac{\sin(\bar\mu_i c_i)}{\bar\mu_i}\\ 
&\frac{1}{\sqrt{p_i}}\rightarrow\frac{3}{8\pi\gamma\ell_P^2}\left(\sqrt{p_i+\frac{8\pi\gamma\ell_P^2}{3}\,\bar\mu_i}-\sqrt{p_i-\frac{8\pi\gamma\ell_P^2}{3}\,\bar\mu_i}\right)\,.\label{invv}
\end{align}
As a consequence, the condition that physical states are annihilated by the scalar constraint becomes a difference equation, which can be solved via numerical tools. 
The values of $\bar\mu$'s are chosen such that the minimum eigenvalue of the physical area operators coincides with that in LQG, so getting the following conditions
\be
\bar\mu_1\bar\mu_2=\frac{\Delta l_P^2}{p_3}\quad
\bar\mu_2\bar\mu_3=\frac{\Delta l_P^2}{p_1}\quad
\bar\mu_3\bar\mu_1=\frac{\Delta l_P^2}{p_2}\,,\label{barmu}
\ee
with $\Delta=4\pi\sqrt{3}\gamma$.

The result of numerical simulations \cite{Singh:2012zc} (see \cite{Ashtekar:2006wn} for sharply peaked states, \cite{Diener:2014mia} and \cite{Diener:2014hba} for wide, squeezed and more general states, and \cite{MartinBenito:2009qu} in the presence of anisotropies) shows how the basic features of the evolution of semiclassical states, in particular the bounce replacing the initial singularity, is already captured by the classical Hamiltonian associated to the operator $\hat{\mathcal{S}}$ in which inverse volume corrections (the kind of corrections coming from \eqref{invv}) are neglected, namely
\begin{align}
H_{eff}=\frac{N}{8\pi G\gamma^2}\bigg(&\sqrt{\frac{p_1p_2}{p_3}}\,\frac{\sin(\bar\mu_1 c_1)}{\bar\mu_1}\,\frac{\sin(\bar\mu_2 c_2)}{\bar\mu_2}+\sqrt{\frac{p_2p_3}{p_1}}\,\frac{\sin(\bar\mu_2 c_2)}{\bar\mu_2}\,\frac{\sin(\bar\mu_3 c_3)}{\bar\mu_3}\nonumber\\
&+\sqrt{\frac{p_3p_1}{p_2}}\,\frac{\sin(\bar\mu_3 c_3)}{\bar\mu_3}\,\frac{\sin(\bar\mu_1 c_1)}{\bar\mu_1}\bigg)=0\,.
\label{effSlqc}
\end{align}
In QRLG, we can infer an effective Hamiltonian of this sort. 

\section{Kinematics of Quantum Reduced Loop Gravity}

The idea of QRLG is to start from the kinematical Hilbert space of LQG and to implement the gauge fixing conditions restricting to a diagonal metric tensor and to diagonal triads. 

Under very general conditions (see for instance \cite{Grant:2009zz}), a generic three-dimensional metric tensor can be diagonalized via a spatial diffeomorphisms, such that one has
\be
dl^2= (a_1(t,x))^2\,dx_1^2+ (a_2(t,x))^2\,dx_2^2+ (a_3(t,x))^2\,dx_3^2\,, \label{dm}
\ee
the three scale factors being generic functions of all spacetime coordinates. In what follows, we will denote $x_1$, $x_2$ and $x_3$ as fiducial coordinates and we will call fiducial directions the corresponding directions in space. 

The gauge-fixing condition giving the restriction to a diagonal metric can be written in terms of inverse densitized triads as
\be
\eta^{ab}=\delta_{ij}\,E^a_i\,E^b_j=0\qquad a\neq b\,.\label{dmgf}
\ee
It is worth noting how we are performing only a partial gauge fixing. In fact, there are some diffeomorphisms which do not generate any off-diagonal component when acting on \eqref{dm} and they can be seen as redefinitions of fiducial coordinates, namely 
\be
x'_1=x_1'(x_1)\quad x_2'=x_2'(x_2)\quad x_3'=x_3'(x_3)\,.\label{reddiff}
\ee
These transformations are residual symmetries of the theory after having fixed \eqref{dmgf}.

A possible set of triads for the metric \eqref{dm} is
\be
e^i_a= a_i\,\delta^i_a\,, \label{dt}
\ee
and the most generic kind of triads are obtained from \eqref{dt} by acting with a $SU(2)$ rotation on the internal index. Hence, the choice of diagonal triads implies breaking manifestly the invariance under internal rotations, whose associated gauge-fixing condition in terms of inverse densitized triads reads
\be
\chi_i=\epsilon_{ij}^{\phantom{12}k}\,E^a_k\,\delta^j_a = 0\,.\label{dtgf}
\ee
This is a complete gauge fixing of the invariance under internal rotation. 

In \cite{Alesci:2013xya} the conditions \eqref{dmgf} and \eqref{dtgf} have been properly smeared, promoted to operators $\hat{\eta}^{ab}$ and $\hat{\chi}_i$ in $\mathcal{H}$ and then implemented weakly, {\it i.e.} we looked for the subspace in which for any two states $|\phi\rangle$ and $|\psi\rangle$ 
\be
\langle \psi | \hat{\eta} | \phi \rangle = 0 \qquad \langle \psi | \hat{\chi} | \phi \rangle = 0\,.
\ee 
These conditions hold if:

\begin{itemize}
{\item the graphs are cuboidal, whose links are only those along one of the fiducial directions.}
{\item the group elements belong to proper $U(1)$ subgroups. In particular, given a link along the fiducial direction $\delta_i=\delta_i^a\partial_a$, the attached $U(1)$ subgroup is obtained by stabilizing the $SU(2)$ group along the internal directions $\vec{u}_i$, where 
\be
\vec{u}_1=(1,0,0)\quad \vec{u}_2=(0,1,0)\quad \vec{u}_3=(0,0,1)\,.
\ee
}
\end{itemize}

The former is easily implemented on $SU(2)$ spin networks \eqref{spinnet solite} by taking as admissible graphs $\Gamma$ only cuboidal ones, thus restricting admissible diffeomorphims to those preserving the cuboidal structure, which are the transformations \eqref{reddiff} \cite{Alesci:2013xd} . The latter is realized at each link by projecting the magnetic indexes of the Wigner matrices on the $SU(2)$ coherent states $|\pm j,  \vec{u}_l \rangle$ having maximum or minimum magnetic number along $\vec{u}_l$ (an interpretation in terms of projected spin networks \cite{Dupuis:2010jn} can be given). The explicit expression of such coherent states in terms of the standard $SU(2)$ basis vectors $|j,m \rangle$ is given by 
\be
|\pm j,  \vec{u}_l \rangle=\sum_{m=-j}^{j}\, R_{l\,\, \pm j m}\,|j,m \rangle\,,
\ee
$R_{l}$ being the rotation mapping the direction $\vec{u}_l$ into $\vec{u}_3$ and its first magnetic index in the expression above has been projected onto $\langle j_{l},\pm j_l |$. We introduce now a convenient graphic notation, in which we denote Wigner matrices by circles and the rotations $R_l$ by squares, such that a generic basis element of the Hilbert space after gauge-fixing reads 
\be
\langle j,n | m,\vec{u}_l\rangle \langle m,\vec{u}_l| D^j(h)| m,\vec{u}_l\rangle \langle m,\vec{u}_l|j,r \rangle
=\begin{array} {c} 
\ifx\JPicScale\undefined\def\JPicScale{1.1}\fi
\psset{unit=\JPicScale mm}
\psset{linewidth=0.3,dotsep=1,hatchwidth=0.3,hatchsep=1.5,shadowsize=1,dimen=middle}
\psset{dotsize=0.7 2.5,dotscale=1 1,fillcolor=black}
\psset{arrowsize=1 2,arrowlength=1,arrowinset=0.25,tbarsize=0.7 5,bracketlength=0.15,rbracketlength=0.15}
\begin{pspicture}(0,0)(68,8)
\psline{|*-}(52,3)(48,3)
\rput(35,8){$\scr{j}$}
\rput{0}(35,3){\psellipse[](0,0)(3,-3)}
\rput(35,3){$h$}
\psline[fillstyle=solid]{-|}(12,3)(16,3)
\pspolygon[](6,6)(12,6)(12,0)(6,0)
\psline[fillstyle=solid](2,3)(6,3)
\rput(9,3){$\scr{R_l}$}
\psline[fillstyle=solid](64,3)(68,3)
\pspolygon[](58,6)(64,6)(64,0)(58,0)
\psline[fillstyle=solid]{|*-}(54,3)(58,3)
\rput(61,3){$\scr{R^{-1}_l}$}
\pspolygon[](42,6)(48,6)(48,0)(42,0)
\rput(45,3){$\scr{R_l}$}
\psline[fillstyle=solid](38,3)(42,3)
\psline{}(32,3)(28,3)
\pspolygon[](22,6)(28,6)(28,0)(22,0)
\rput(25,3){$\scr{R^{-1}_l}$}
\psline[fillstyle=solid]{|-}(18,3)(22,3)
\rput(9,8){$\scr{j}$}
\rput(61,8){$\scr{j}$}
\end{pspicture}
\end{array}\qquad m=\pm j\,,
\label{Dridottamn}
\ee
where the breaks are the projections on the maximum or minimum magnetic numbers, {\it i.e.} $|j,\pm j\rangle \langle j,\pm j |$. It is worth noting how there are two rotations $R^{-1}_l$ and $R_l$ before and two rotations $R_l$ and $R_l^{-1}$ after the $SU(2)$ group element. Among them, those immediately before and after the group element combine with the Wigner matrix to form the representation of the $U(1)$ subgroup along the link $l$, which we call $ \;{}^l\!D^{j_{l}}_{m_{l} m_{l}}(h_{l})$ for $m_l=\pm j_l$, while the additional rotations attach to the intertwiners, such that a generic basis element can be written as 
\be
\langle h|\Gamma, {\bf m_l, x_\vgraph \bf}\rangle= \prod_{n\in\Gamma}\langle{\bf j_{l}}, {\bf x}_n|{\bf m_{l}},  \vec{{\bf u}}_l \rangle 
\prod_{l} \;{}^l\!D^{j_{l}}_{m_{l} m_{l}}(h_{l}),\quad m_l=\pm j_l
\label{base finale}
\ee
where the products $\prod_{n\in\Gamma}$ and $\prod_{l}$ extend over all the nodes $n\in\Gamma$ and over all the links $l$ emanating from $n$. One sees how some nontrivial coefficients $\langle{\bf j_{l}}, {\bf x}_n|{\bf m_{l}},  \vec{{\bf u}}_l \rangle$ are now present at nodes: they are some one-dimensional intertwiners proper of QRLG (they are the projection of Livine-Speziale coherent intertwiners \cite{Livine:2007vk} on the standard $SU(2)$ intertwiners basis). The mathematical reason for the appearance of such intertwiners is that the $U(1)$ subgroups along different fiducial directions are not independent, because they are obtained by stabilizing the same $SU(2)$ group along different internal directions, such that the associated $U(1)$ representations have nonvanishing projections among each others. The presence of intertwiners is the main technical achievement of QRLG. 

The action of the flux operators is given by taking those in LQG and projecting down to QRLG as explained before for states. As a consequence, fluxes $E_i(S^j)$ are nonvanishing only when $i=j$ and they behave as the invariant vector fields associated with the corresponding $U(1)$ subgroups, {\it i.e.} 
\be
\hat{E}_i(S^i){}^l\!D^{j_{l}}_{m_{l} m_{l}}(h_{l})= 8\pi\gamma l_P^2\, m_{l}\,{}^l\!D^{j_{l}}_{m_{l} m_{l}}(h_{l}) \qquad l_i\cap S^i\neq \oslash .\label{redei}
\ee
The composite operators we can construct out of fluxes are extremely simple. For instance, the volume operator $\hat{V}=\int d^3x \sqrt{|\hat{E}_1(S^1)\hat{E}_2(S^2)\hat{E}_3(S^3)|}$ is diagonal. 

Finally, to compute the action of holonomy operators, we need the recoupling theory, which is just that of the $U(1)$ group, {\it i.e.}
\be
 \begin {split}
\begin{array}{c}
\ifx\JPicScale\undefined\def\JPicScale{1}\fi
\psset{unit=\JPicScale mm}
\psset{linewidth=0.3,dotsep=1,hatchwidth=0.3,hatchsep=1.5,shadowsize=1,dimen=middle}
\psset{dotsize=0.7 2.5,dotscale=1 1,fillcolor=black}
\psset{arrowsize=1 2,arrowlength=1,arrowinset=0.25,tbarsize=0.7 5,bracketlength=0.15,rbracketlength=0.15}
\begin{pspicture}(0,0)(42,19)
\psline(19,8)(19,0)
\psline{-<<}(19,4)(12,4)
\psline{<-}(32,4)(25,4)
\psline(19,8)(25,4)
\psline(25,4)(19,0)
\psline(19,19)(19,11)
\psline{-<<}(19,15)(12,15)
\psline{<-}(32,15)(25,15)
\psline(19,19)(25,15)
\psline(25,15)(19,11)
\rput(29,2){$j_1$}
\rput(28,13){$j_2$}
\psline{<-|}(42,4)(35,4)
\psline{<-|}(42,15)(35,15)
\psline{|-<<}(9,4)(2,4)
\psline{|-<<}(9,15)(2,15)
\psline(12,16)(12,14)
\psline(32,16)(32,14)
\psline(32,5)(32,3)
\psline(12,5)(12,3)
\rput(6,2){$j_1$}
\rput(6,13){$j_2$}
\rput(38,13){$j_2$}
\rput(38,2){$j_1$}
\end{pspicture}
\end{array}
=
\begin{array}{c}
\ifx\JPicScale\undefined\def\JPicScale{1}\fi
\psset{unit=\JPicScale mm}
\psset{linewidth=0.3,dotsep=1,hatchwidth=0.3,hatchsep=1.5,shadowsize=1,dimen=middle}
\psset{dotsize=0.7 2.5,dotscale=1 1,fillcolor=black}
\psset{arrowsize=1 2,arrowlength=1,arrowinset=0.25,tbarsize=0.7 5,bracketlength=0.15,rbracketlength=0.15}
\begin{pspicture}(0,0)(42,12)
\psline(19,10)(19,2)
\psline{-<<}(19,6)(12,6)
\psline{<-}(32,6)(25,6)
\psline(19,10)(25,6)
\psline(25,6)(19,2)
\rput(28,4){$\scr{|n_1+n_2|}$}
\rput(36,6){$\scr{n_1+n_2}$}
\psline{<-|}(42,1)(35,1)
\psline{<-|}(42,12)(35,12)
\psline{|-<<}(9,1)(2,1)
\psline{|-<<}(9,12)(2,12)
\psline(12,7)(12,5)
\psline(32,7)(32,5)
\rput(6,-1){$j_1$}
\rput(6,10){$j_2$}
\rput(38,10){$j_2$}
\rput(38,-1){$j_1$}
\rput(11,12){$\scr{n_2}$}
\rput(11,2){$\scr{n_1}$}
\rput(33,12){$\scr{n_2}$}
\rput(32,1){$\scr{n_1}$}
\end{pspicture}
\end{array}
\end{split}\label{newrec}
 \ee 

The physical implications of this choice are discussed in \cite{Alesci:2015nja} .

\section{Dynamics of Quantum Reduced Loop Gravity}

The formulation we have been considering till now is a gauge-fixed LQG, which means that we did not perform any reduction of degrees of freedom. We make such a reduction now, on a dynamical level, by considering only that part of the scalar constraint which generates the evolution of the homogeneous part of the metric. 

In the homogeneous limit, the scale factors are just functions of time. Indeed, one can retain the invariance under reduced diffeomorphisms requiring that
\be
a_i=a_i(t,x_i)\,,
\ee
which means that the metric tensor is homogeneous up to a redefinition of fiducial coordinates. Once this condition holds it can be shown that spin connections vanish, so Ashtekar-Barbero connections are diagonal. Furthermore, the vector constraint vanishes identically, while the $SU(2)$ Gauss constraint reduces to three 
independent $U(1)$ Gauss constraints along each principal direction. The full dynamics is obtained by substituting the diagonal form for connections and momenta into the scalar constraint of full theory. As a consequence, the Lorentzian part is proportional to the Euclidean part and one obtains \eqref{S}. Hence, when we refer to the part of the scalar constraint which generates the evolution of the homogeneous part, we refer to the simplifications occurring when homogeneity holds, {\it i.e.} the scalar constraint is proportional to the Euclidean part. 

If we allow $a_i$ to be generic functions of all fiducial coordinates, then neither Ashtekar-Barbero connections are diagonal anymore, neither the vector constraint vanishes identically. The Hamiltonian analysis in reduced phase space is so complicated that it is really hard to try to figure out how to quantize the resulting system (see for instance \cite{Bodendorfer:2014vea} ). A different way to tackle the problem is to consider a perturbative expansion around a homogeneous configuration and what we are going to present is the leading order term of such an expansion. This is enough for cosmology (or in the limit in which the Belinski Lipschitz Kalatnikov conjecture works \cite{bkl1,Belinsky:1982pk} , when we can neglect spatial gradients with respect to time derivatives), while the next-to-the-leading order terms will be discussed in forthcoming papers with the aim to characterize the behavior of perturbations and to extract phenomenological implications.

Hence, we take the Euclidean part of the scalar constraint as the full Hamiltonian (modulo a coefficient) and we replace the operators in LQG with the corresponding expressions in QRLG. Hence, the dynamics is generated by the following operator
\be\label{hamEc}
\hat{H}=\frac{1}{16\pi G\gamma^2}\sum_{\vgraph} \hat{H}^\vgraph_{E}[N]=\frac{1}{16\pi G\gamma^2}\sum_{\vgraph}\sum_{\cube} \hat{H}^\vgraph_{E\cube}[N], 
\ee
where the summations extend over all the nodes $\vgraph$ of the cuboidal graph at which the states are based and over all the triples of links emanating from the same node, and the action of $\hat{H}^\vgraph_{E\cube}[N]$ on a n-valent node reads
\be
\hat{H}^\vgraph_{E\cube}[N]:= \frac{4i}{8\pi \gamma l^2_P\,c(n)}\mathcal{N}(\vgraph)  \, \epsilon^{ijk} \,
   \mathrm{Tr}\Big[\hat{h}_{\alpha_{ij}} \hat{h}^{-1}_{s_{k}} \big[\hat{h}_{s_{k}},\hat{V}\big]\Big]\qquad c(n)=2^{n-3}\,, 
   \label{Hridotto}
\ee

$\mathcal{N}(\vgraph)$ being the lapse function in $\vgraph$, while the trace denotes the sum over the two possible values of the magnetic indexes for the considered holonomies. The difference with respect to the expression \eqref{Hm_delta:quantum} is that we take the holonomies in the fundamental representation and that we consider a regularization based on a cubulation rather than a triangulation. The latter is also partially responsible for the fact that a different normalization factor is present with respect to \cite{Thiemann:1996aw,Gaul:2000ba,io e antonia,Alesci:2013kpa} , while the coefficient $c(n)$ is the total number of non-coplanar triples, which are the only triples contributing to \eqref{Hridotto}.

Since the volume operator is diagonal, we can analytically compute the action of \eqref{hamEc} on the states of QRLG. In what follows, we consider a non-graph changing Hamiltonian, which means that the Hamiltonian adds some links already present in the original graph. 

In particular, let us consider the most generic case of a state based at a graph having six-valent nodes. We use the following notation: we write $j_l=j^{(i)}_\vgraph$, the superscript $^{(i)}$ being the fiducial direction of the link and the subscript $_\vgraph$ the base point, while we introduce the space vectors
$\vec{e}_{i}$ along the fiducial direction $i$ connecting two first neighbor nodes, such that $j^{(i)}_{\vgraph-\vec{e}_j}$ is the spin number of the link along $i$ starting in the node $\vgraph-\vec{e}_j$. Hence, at a given node for a fixed triple of links we have the following action \footnote{We present only the case with positive $m_l=j_l$, the extension to $m_l=-j_l$ being straightforward.} 

\be
\begin{split}
&\mathrm{Tr}\Big[\hat{h}_{\alpha_{12}} \hat{h}^{-1}_{s_3} \!\hat{V} \hat{h}_{s_{3}}\Big] |\Gamma, {\bf j_l, x_\vgraph \bf}\rangle
=\\
&=(8\pi\gamma l_P^2)^{3/2}\sum_{\mu'_1,\mu'_2,\mu_2,\mu_1=\pm \frac{1}{2}}\sum_{\mu=\pm \frac{1}{2}} \sqrt{j^{(1)}_{\vgraph}\;j^{(2)}_{\vgraph}\;(j^{(3)}_{\vgraph}+\mu)}\; s(\mu)  C^{1\,0}_{\frac{1}{2}\,\frac{1}{2} \; \frac{1}{2}\,-\frac{1}{2}}\\
&\ba
\ifx\JPicScale\undefined\def\JPicScale{1}\fi
\psset{unit=\JPicScale mm}
\psset{linewidth=0.3,dotsep=1,hatchwidth=0.3,hatchsep=1.5,shadowsize=1,dimen=middle}
\psset{dotsize=0.7 2.5,dotscale=1 1,fillcolor=black}
\psset{arrowsize=1 2,arrowlength=1,arrowinset=0.25,tbarsize=0.7 5,bracketlength=0.15,rbracketlength=0.15}
\begin{pspicture}(0,0)(173,90)
\psline[fillstyle=solid]{-|}(68.83,70.83)(66,68)
\pspolygon[](72,72)(70,70)(68,72)(70,74)
\rput(66,73){$\scr{R_1}$}
\rput(71,76){$\scr{j_{\vgraph}^{(1)}}$}
\psline[fillstyle=solid]{-|}(77,79)(77,89)
\psline[fillstyle=solid]{-|}(89,79)(93,79)
\rput(89.46,82.54){$\scr{R_2}$}
\rput(83.81,81.12){$\scr{j_{\vgraph}^{(2)}}$}
\rput(74,63){$\scr{R_1}$}
\rput(82,70){$\scr{1/2}$}
\psline[fillstyle=solid]{|*-}(93,73)(89,73)
\rput(91.59,70.44){$\scr{R_2^{-1}}$}
\rput(81,76){$\scr{1}$}
\pspolygon[](88.76,71.86)(85.93,71.86)(85.93,74.69)(88.76,74.69)
\psline[fillstyle=solid](86,73)(79,73)
\rput(73,89){$\scr{j_{\vgraph}^{(3)}}$}
\psline[fillstyle=solid](77,79)(70.84,72.84)
\pspolygon[](85.93,80.41)(88.76,80.41)(88.76,77.59)(85.93,77.59)
\rput(79,86){$\scr{0}$}
\psline[fillstyle=solid](77.51,79)(86,79)
\psline(77,79)(77,65)
\psline[border=0.3,fillstyle=solid](75,69)(79,73)
\psline[fillstyle=solid]{-|}(66,79)(62,79)
\pspolygon[](69.07,77.59)(66.24,77.59)(66.24,80.41)(69.07,80.41)
\psline[fillstyle=solid](77.49,79)(69,79)
\psline[fillstyle=solid]{-|}(85.17,87.17)(88,90)
\pspolygon[](82,86)(84,88)(86,86)(84,84)
\psline[fillstyle=solid](77,79)(83.16,85.16)
\psline[border=0.3,fillstyle=solid](79,73)(79,84)
\psline[fillstyle=solid]{-|}(18.83,20.83)(16,18)
\pspolygon[](22,22)(20,20)(18,22)(20,24)
\psline[fillstyle=solid]{-|}(27,29)(27,39)
\psline[fillstyle=solid]{-|}(39,29)(43,29)
\rput(39.46,32.54){$\scr{R_2}$}
\rput(21,38){$\scr{j^{(3)}_{\vgraph+\vec{e}_1}}$}
\psline[fillstyle=solid](27,29)(20.84,22.84)
\pspolygon[](35.93,30.41)(38.76,30.41)(38.76,27.59)(35.93,27.59)
\psline[fillstyle=solid](27.51,29)(36,29)
\psline(27,29)(27,15)
\psline[fillstyle=solid]{-|}(16,29)(12,29)
\pspolygon[](19.07,27.59)(16.24,27.59)(16.24,30.41)(19.07,30.41)
\psline[fillstyle=solid](27.49,29)(19,29)
\psline[fillstyle=solid]{-|}(35.17,37.17)(38,40)
\pspolygon[](32,36)(34,38)(36,36)(34,34)
\rput(34,41){$\scr{R^{-1}_1}$}
\rput(39,38){$\scr{j_{\vgraph}^{(1)}}$}
\psline[fillstyle=solid](27,29)(33.16,35.16)
\psline[fillstyle=solid]{|*-}(70.17,64.17)(73,67)
\pspolygon[](72,68)(74,70)(76,68)(74,66)
\psline{|*-}(66,64)(63.17,61.17)
\rput{45}(56.5,54.5){\psellipse[](0,0)(2.21,-2.12)}
\rput(61,55){$\scr{h_{1}}$}
\rput(66,59){$\scr{R^{-1}_1}$}
\psline[fillstyle=solid](58.17,56.17)(61,59)
\psline(55,53)(52.18,50.18)
\pspolygon[](49,49)(51,51)(53,49)(51,47)
\rput(55,49){$\scr{R_1}$}
\psline[fillstyle=solid]{|-}(47,45)(49.83,47.83)
\pspolygon[](60,60)(62,62)(64,60)(62,58)
\rput(51,59){$\scr{j_{\vgraph}^{(1)}+\mu_1}$}
\psline[fillstyle=solid]{|*-}(46.83,40.83)(44,38)
\rput(48,39){$\scr{R^{-1}_1}$}
\rput(53,42){$\scr{\mu_1}$}
\pspolygon[](45,37)(43,35)(41,37)(43,39)
\psline[fillstyle=solid]{-|}(47.34,33.93)(51.34,33.93)
\pspolygon[](44.27,35.34)(47.1,35.34)(47.1,32.51)(44.27,32.51)
\rput(49,36){$\scr{R_2}$}
\psline[fillstyle=solid](40.03,33.93)(44.5,33.93)
\psline[fillstyle=solid](42,36)(40,34)
\psline{|*-}(123.88,75.88)(119.88,75.88)
\rput(109.03,69.51){$\scr{j_{\vgraph}^{(2)}-\mu_2}$}
\rput(109.74,80.12){$\scr{h_{2}}$}
\rput(115.39,78.71){$\scr{R_2}$}
\psline[fillstyle=solid](112.81,75.88)(116.81,75.88)
\psline(108.32,75.88)(104.33,75.88)
\pspolygon[](101.25,77.29)(104.08,77.29)(104.08,74.46)(101.25,74.46)
\rput(103.37,80.83){$\scr{R^{-1}_2}$}
\psline[fillstyle=solid]{|-}(97.01,75.88)(101.01,75.88)
\pspolygon[](116.81,77.29)(119.64,77.29)(119.64,74.46)(116.81,74.46)
\rput{90}(110.44,75.79){\psellipse[](0,0)(2.21,-2.12)}
\psline{|*-}(83.88,31.88)(79.88,31.88)
\rput(69.74,36.12){$\scr{h_{2}}$}
\rput(75.39,34.71){$\scr{R_2}$}
\psline[fillstyle=solid](72.81,31.88)(76.81,31.88)
\psline(68.32,31.88)(64.33,31.88)
\pspolygon[](61.25,33.29)(64.08,33.29)(64.08,30.46)(61.25,30.46)
\rput(63,36){$\scr{R^{-1}_2}$}
\psline[fillstyle=solid]{|-}(57,32)(61,32)
\pspolygon[](76.81,33.29)(79.64,33.29)(79.64,30.46)(76.81,30.46)
\rput{90}(70.44,31.79){\psellipse[](0,0)(2.21,-2.12)}
\psline[fillstyle=solid]{-|}(148.83,69.83)(146,67)
\pspolygon[](152,71)(150,69)(148,71)(150,73)
\rput(146,72){$\scr{R_1}$}
\rput(151,66){$\scr{j^{(1)}_{\vgraph+\vec{e}_2}}$}
\psline[fillstyle=solid]{-|}(157,78)(157,88)
\psline[fillstyle=solid]{-|}(169,78)(173,78)
\psline[fillstyle=solid](157,78)(150.84,71.84)
\pspolygon[](165.93,79.41)(168.76,79.41)(168.76,76.59)(165.93,76.59)
\psline[fillstyle=solid](157.51,78)(166,78)
\psline(157,78)(157,64)
\rput(144,82){$\scr{R^{-1}_2}$}
\rput(151,81){$\scr{j_{\vgraph}^{(2)}}$}
\psline[fillstyle=solid]{-|}(146,78)(142,78)
\pspolygon[](149.07,76.59)(146.24,76.59)(146.24,79.41)(149.07,79.41)
\psline[fillstyle=solid](157.49,78)(149,78)
\pspolygon[](162,85)(164,87)(166,85)(164,83)
\psline[fillstyle=solid](157,78)(163.16,84.16)
\psline[fillstyle=solid]{-|}(98.83,20.83)(96,18)
\pspolygon[](102,22)(100,20)(98,22)(100,24)
\psline[fillstyle=solid](107,29)(100.84,22.84)
\psline(107,29)(107,15)
\rput(95,32){$\scr{R^{-1}_2}$}
\psline[fillstyle=solid]{-|}(96,29)(92,29)
\pspolygon[](99.07,27.59)(96.24,27.59)(96.24,30.41)(99.07,30.41)
\psline[fillstyle=solid](107.49,29)(99,29)
\psline[fillstyle=solid]{-|}(115.17,37.17)(118,40)
\pspolygon[](112,36)(114,38)(116,36)(114,34)
\rput(118,35){$\scr{R^{-1}_1}$}
\psline[fillstyle=solid](107,29)(113.16,35.16)
\psline{|*-}(138,64)(135.17,61.17)
\rput{45}(128.5,54.5){\psellipse[](0,0)(2.21,-2.12)}
\rput(133,55){$\scr{h_{1}}$}
\rput(130,61){$\scr{R^{-1}_1}$}
\psline[fillstyle=solid](130.17,56.17)(133,59)
\psline(127,53)(124.18,50.18)
\pspolygon[](121,49)(123,51)(125,49)(123,47)
\rput(117,49){$\scr{R_1}$}
\psline[fillstyle=solid]{|-}(119,45)(121.83,47.83)
\pspolygon[](132,60)(134,62)(136,60)(134,58)
\rput(106,42){$\scr{R_1}$}
\psline[fillstyle=solid]{|*-}(92,35)(96,35)
\rput(93,39){$\scr{R_2^{-1}}$}
\pspolygon[](96.24,36.14)(99.07,36.14)(99.07,33.31)(96.24,33.31)
\psline[fillstyle=solid](99,35)(106,35)
\psline[fillstyle=solid](110,39)(106,35)
\psline[fillstyle=solid]{|*-}(114.83,43.83)(112,41)
\pspolygon[](113,40)(111,38)(109,40)(111,42)
\psline[fillstyle=solid]{|*-}(138.17,67.17)(141,70)
\rput(137,69){$\scr{R^{-1}_1}$}
\pspolygon[](140,71)(142,73)(144,71)(142,69)
\psline[fillstyle=solid]{-|}(137.66,74.07)(133.66,74.07)
\pspolygon[](140.73,72.66)(137.9,72.66)(137.9,75.49)(140.73,75.49)
\rput(136,77){$\scr{R_2}$}
\psline[fillstyle=solid](144.97,74.07)(140.5,74.07)
\psline[fillstyle=solid](143,72)(145,74)
\psline[border=0.3,fillstyle=solid]{-|}(107,29)(107,39)
\psline[fillstyle=solid]{-|}(119,29)(123,29)
\pspolygon[](115.93,30.41)(118.76,30.41)(118.76,27.59)(115.93,27.59)
\psline[fillstyle=solid](107.51,29)(116,29)
\psline[fillstyle=solid]{-|}(165.17,86.17)(168,89)
\rput(33,26){$\scr{j^{(2)}_{\vgraph+\vec{e}_1}}$}
\rput(69,27){$\scr{j^{(2)}_{\vgraph+\vec{e}_1}+\mu'_2}$}
\rput(93,26){$\scr{j^{(2)}_{\vgraph+\vec{e}_1}}$}
\rput(122,38){$\scr{j^{(1)}_{\vgraph+\vec{e}_2}}$}
\rput(116,54){$\scr{j^{(1)}_{\vgraph+\vec{e}_2}-\mu'_1}$}
\rput(70,62){$\scr{\mu_1}$}
\rput(53,34){$\scr{\mu'_2}$}
\rput(90,35){$\scr{\mu'_2}$}
\rput(114,46){$\scr{\mu'_1}$}
\rput(136,67){$\scr{\mu'_1}$}
\rput(131,74){$\scr{\mu_2}$}
\rput(94,72){$\scr{\mu_2}$}
\end{pspicture}
\ea\,,
\end{split}
\ee
where we notice how the quantum numbers at links have changed, while the new intetrtwiners are the product of the original ones with those given by the operator. On the right-hand side of the expression above, the state is not in the proper form, since we should rewrite the products of the two intertwiners at the nodes in terms of the new intertwiners, corresponding to the new quantum numbers after the action of the considered operator. This can always be done, since the new intertwiners are nonvanishing provided that the original one are not vanishing and we can multiply and divide the expression above times the new intertwiners, so getting the state with the right intertwiner times a coefficient. For the sake of the semiclassical analysis, this computation is unnecessary.

In order to get the action of the full Hamiltonian, this result has then to be summed over all the possible permutations of the links within the chosen triple, then over all triples for a given nodes and, finally, over all the nodes of the graph.

\subsection{Semiclassical analysis} 

Semiclassical states can be defined using the tools developed for full LQG \cite{Thiemann:2000bw,Thiemann:2002vj} as follows \cite{Alesci:2014uha}
\begin{equation}
\psi^{{\bf\alpha}}_{\Gamma{\bf H'}}=\sum_{{\bf m_{l}}}\prod_{n\in\Gamma} \langle{\bf j_{l}}, {\bf x}_n|{\bf m_{l}},  \vec{{\bf u}}_l \rangle^*\;\prod_{l\in\Gamma} \psi^\alpha_{H'_{l}}(m_{l})\;\langle h|\Gamma, {\bf m_l, x_\vgraph \bf}\rangle\,,
\label{semiclassici ridotti inv}
\end{equation}
where the coefficients $\psi^\alpha_{H'}(m_{l})$ are given by
\begin{equation}
\psi^\alpha_{H'}(m_{l})=(2j_{l}+1)e^{-j_{l}(j_{l}+1)\frac{\alpha}{2}}e^{i\theta_lm_{l}}e^{\frac{\alpha}{8\pi\gamma l_P^2}E'_im_{l}}\,,\quad j_l=|m_l|\,.
\end{equation}
and they provide at each link the requested peakedness properties around a classical configuration having dual fluxes $E'_i=8\pi\gamma \ell_P^2 \bar{j}_i$\footnote{We choose to peak around positive values of $m$'s, such that $m_l=j_l$.} and holonomy $h'=e^{i\theta_l}$. We assume to peak around a homogeneous configuration, such that $\bar{j}_l=\bar{j}^{(i)}_\vgraph=\bar{j}_i$ and $\theta_l=\theta^{(i)}_\vgraph=\theta_i$, based at a graph having $N$ six-valent nodes. We denote the associated semiclassical state by $|\Psi_{N,H}\rangle$. Since $E'_i$ and $\theta_l$ are single-cell variables, in terms of LQC phase space variables we have the following identifications
\be
E'_i=p_i\,\frac{N_i}{N} \qquad \theta_l=\frac{c_l}{N_l}\,,\label{lqcvar}
\ee
where we introduce the number of nodes $N_i$ along the direction $i$ and $N=N_1N_2N_3$.
At the leading order of the semiclassical expansion, we get the following expectation value for the operator \eqref{Hridotto}
\be
\langle\Psi_{N,H} |\hat{H}^\vgraph_{E\cube} |\Psi_{N,H} \rangle\approx
2\mathcal{N}(\vgraph)(8\pi\gamma l_P^2)^{1/2}\sum_{\mu=\pm 1/2}\sqrt{\bar{j}_1\;\bar{j}_2\;(\bar{j}_3+\mu)}\; s(\mu) \sin{\theta_1} \sin{\theta_2},
\ee 
where we took the loop $\alpha_{ij}$ in the plane $12$ and $s(\mu)$ denotes the sign function. By taking a large $j$ expansion, we can write
\be
\sum_{\mu=\pm 1/2}\sqrt{\bar{j}_3+\mu}\; s(\mu)\approx \frac{1}{\bar{j}_3} \left[1+\frac{1}{(8\bar{j}_3)^2}\right]\,,\label{ivexp}
\ee
and retaining only the leading contribution we get
\be
\langle\Psi_{N,H}|\hat{H}^\vgraph_{E\cube} |\Psi_{N,H} \rangle\approx
2(8\pi\gamma l_P^2)^{1/2}\mathcal{N}(\vgraph)\sqrt{\frac{\bar{j}_1\;\bar{j}_2}{\bar{j}_3}}\;  \sin{\theta_1} \sin{\theta_2}\,.\label{1ham}
\ee 

By summing over all the triples and nodes we obtain 
\be
\langle\Psi_{N,H}|\sum_{\vgraph,\cube}\hat{H}^\vgraph_{E\cube} |\Psi_{N,H}\rangle\approx
4N(8\pi\gamma l_P^2)^{1/2} \mathcal{N}(\vgraph)\bigg(\sqrt{\frac{\bar{j}_1\;\bar{j}_2}{\bar{j}_3}}\;  \sin{\theta_1} \sin{\theta_2} + \sqrt{\frac{\bar{j}_2\;\bar{j}_3}{\bar{j}_1}}\;  \sin{\theta_2} \sin{\theta_3} + \sqrt{\frac{\bar{j}_3\;\bar{j}_1}{\bar{j}_2}}\;  \sin{\theta_3} \sin{\theta_1}\bigg)\,.\label{ham6}
\ee 
If we rewrite the expression above in terms of LQC variables \eqref{lqcvar}, we get for the expectation value of the full Hamiltonian \eqref{hamEc}
\begin{align}
\langle\Psi_{N,H}|& \hat{H} |\Psi_{N,H}\rangle
\approx\nonumber\\
&\frac{2}{8\pi G\gamma^2}\mathcal{N}\bigg(N_1\, N_2\,\sqrt{\frac{p_1\;p_2}{p_3}}\;  \sin{\frac{c_1}{N_1}} \sin{\frac{c_2}{N_2}}
+N_2\, N_3\,\sqrt{\frac{p_2\;p_3}{p_1}}\;  \sin{\frac{c_2}{N_2}} \sin{\frac{c_3}{N_3}}
+N_3\, N_1\,\sqrt{\frac{p_3\;p_1}{p_2}}\;  \sin{\frac{c_3}{N_3}} \sin{\frac{c_1}{N_1}}\bigg)\,,\label{Nham}
\end{align}
which coincides with the expression \eqref{effSlqc} if the following identification holds
\be
\bar\mu_i=\frac{1}{N_i}\,.
\ee
Therefore, \emph{in QRLG we effectively obtain the same semiclassical dynamics as in LQC as soon as we identify the regulator with the inverse number of nodes of the graph at which the states are based}. 

Indeed, in this framework the number of node is fixed, since we worked with a non-graph changing Hamiltonian. Hence, we consistently derive the so-called $\mu_0$ scheme, in which the regulator is constant. If we allow the number of nodes to be a function of phase space variables (for instance by adopting a graph-changing Hamiltonian), then the improved scheme \eqref{barmu} is obtained for constant spin numbers $\bar{j}_i$ \cite{Alesci:2014rra} . 

Inverse volume corrections come from the expansion \eqref{ivexp} and by rewriting them in terms of LQC variables we get
\be
\frac{1}{\sqrt{p_i}}\rightarrow \frac{1}{\sqrt{p_i}}\left[1+\frac{N^2}{N^2_i}\,\left(\frac{\pi\gamma l_P^2}{p_i}\right)^2\right]\,.
\ee
The factor $N$ in the expression above implies an enhancement with respect to analogous corrections in LQC \eqref{invv} (see \cite{Alesci:2014rra} for details).

\section{Scalar Field}

Matter fields in LQG have a well-defined description \cite{Thiemann:1997rt,Thiemann:1997rq} , but they are difficult to handle due to the same technical difficulties which affect the formulation in vacuum. These difficulties can be solved in QRLG and in what follows we present the case with scalar fields. 

Let us consider a scalar field $\phi$ in GR, the total scalar constraint is the sum of that of gravity plus that of the scalar field, which can be written as the sum of three terms 
\begin{equation}\label{Hamiltonianconstraint}
\mathcal{S}^{(\phi)}=H^{(\phi)}_{kin}+H^{(\phi)}_{der}+H^{(\phi)}_{pot}\,,
\end{equation}
which are the kinetic, derivative and potential parts and read
\begin{align}
&H^{(\phi)}_{kin}=\frac{\lambda}{2\sqrt{q}}\pi^2\label{kinetic}\\
&H^{(\phi)}_{der}=\frac{\sqrt{q}}{2\lambda}q^{ab}\partial_a\phi\partial_b\phi\label{derivative}\\
&H^{(\phi)}_{pot}=\frac{\sqrt{q}}{2\lambda}V(\phi),,\label{potential}
\end{align}
$\pi$ being the conjugate momentum to the scalar field.

The total Hilbert space is the direct product of that for gravity times that for the scalar field, 
the latter being (see also \cite{Ashtekar:2002sn,Ashtekar:2002vh,Kaminski:2005nc,Kaminski:2006ta} )
\begin{equation}
\mathcal{H}^{(\phi)}:=\overline{\big\{a_1U_{\pi_1}+...+a_nU_{\pi_n}\!:\ a_i\in\mathbb{C},\, n\in\mathbb{N}\big\}}.
\end{equation}
The polymer variable $\langle \phi| U_\pi\rangle=U_\pi(\phi)$ reads
\begin{equation}
U_\pi(\phi)=e^{i\sum_{\vgraph \in\Sigma}\pi_\vgraph \phi_\vgraph}:=\langle{\phi}\ket{U_\pi} \,,
\end{equation}
$\pi$ being a function of finite support given by a countable set of points $\vgraph$, while $\phi_\vgraph=\{\phi(\vgraph_1),\ldots,\phi(\vgraph_n)\}$ is the value of the field $\phi$ at these points. The scalar product is given by
\begin{equation}
\langle U_{\pi} |  U_{\pi'} \rangle:=\delta_{\pi,\pi'}.
\end{equation}
and the basic variables act as follows:
\begin{equation}\label{canonicaloperators-1}
\langle{\phi}|\hat{U}_\pi\ket{U_{\pi'}}=\langle{\phi}\ket{U_{\pi+\pi'}}=e^{i\left(\sum_{\vgraph}\pi_\vgraph \phi_\vgraph + \sum_{\vgraph'}\pi'_{\vgraph'}\phi_{\vgraph'}\right)}, \ \ 
\hat{\Pi}(V)\ket{U_{\pi}}=\hbar\sum_{\vgraph\in V}\pi_\vgraph\ket{U_{\pi}}\,,
\end{equation}
$\Pi(V)$ being the scalar field momentum smeared over the volume $V\subseteq\Sigma$ and $\vgraph\in V$ is the subset of points contained in $V$. 
It is worth noting how the operator associated to $\phi$ does not exist.

The countable set of points at which the scalar field operators are defined coincide with the set of nodes of the graph $\Gamma$ at which the states of QRLG are based.

The quantization of the scalar constraint \eqref{Hamiltonianconstraint} can be done as for gravity by taking the expression in full theory \cite{Thiemann:1997rt} and replacing the holonomies and fluxes of LQG with those of QRLG. For instance, the smeared operator corresponding to the kinetic part \eqref{kinetic} reads
\begin{equation}\label{kin2bis}
\begin{split}
\hat{H}^{(\phi)}_{kin}&[\mathcal{N}]\,\,|\Gamma, {\bf j_l, x_\vgraph , U_{\pi} \bf}\rangle=
-\frac{2^{21}\lambda}{3^2(16\pi\gamma G\hbar)^6}\nonumber\\
&\sum_{\vgraph\in\Gamma} \mathcal{N}(\vgraph)\hat{\Pi}^2(\vgraph)\,
\Bigg[\sum_{\cube}\epsilon_{ijk}\epsilon_{pqr}\,
\text{tr}\bigg(\!
\tau^i\hat{h}_{p}^{-1}\hat{V}^{\!\frac{1}{2}}\hat{h}_{p}
\!\bigg)\,\text{tr}\bigg(\!
\tau^j\hat{h}_{q}^{-1}\hat{V}^{\!\frac{1}{2}}\hat{h}_{q}
\!\bigg)\,\text{tr}\bigg(\!
\tau^k\hat{h}_{r}^{-1}\hat{V}^{\!\frac{1}{2}}\hat{h}_{r}
\!\bigg)\Bigg]^2
\!|\Gamma, {\bf j_l, x_\vgraph , U_{\pi} \bf}\rangle,\!\!
\end{split}
\end{equation}
where $\tau_i$ are $SU(2)$ generators. 
The action of this operator can be explicitly computed giving
\begin{equation}\label{kin5}
\begin{split}
\!\!\hat{H}^{(\phi)}_{kin}[\mathcal{N}]\!|\Gamma, {\bf j_l, x_\vgraph , U_{\pi} \bf}\rangle=&\,
\frac{2^{11}\lambda}{(8\pi\gamma l_P^2)^{\frac{3}{2}}}
\!\sum_{\vgraph}\!\mathcal{N}_{\vgraph}\,\hat{\Pi}^2_{\vgraph}\,\,
\Sigma^{(1)}_\vgraph \Sigma^{(2)}_\vgraph \Sigma^{(3)}_\vgraph\,\, 
\Big(\Delta^{(1),\frac{1}{4}}_\vgraph \Delta^{(2),\frac{1}{4}}_\vgraph \Delta^{(3),\frac{1}{4}}_\vgraph
\Big)^{\!2}
\!|\Gamma, {\bf j_l, x_\vgraph , U_{\pi} \bf}\rangle\,,
\end{split}
\end{equation}
with
\begin{equation}
\begin{split}
\Sigma^{(q)}_{\vgraph}&=\frac{1}{2}(j^{(q)}_{\vgraph}+j^{(q)}_{\vgraph-\vec{e}_q})\\
\Delta^{(p),n}_\vgraph&=\frac{1}{2^n}\left[\Big(\Big|j^{(p)}_\vgraph-1\Big|+j^{(p)}_{\vgraph-\vec{e}_p}\Big)^{n}-
\Big(\Big|j^{(p)}_\vgraph+1\Big|+j^{(p)}_{\vgraph-\vec{e}_p}\Big)^{n}\right]\,.
\end{split}
\end{equation}

Similar expressions can be written for the derivative \eqref{derivative} and potential \eqref{potential} parts \cite{Bilski:2015dra} , so getting the full operator describing the dynamics of the scalar field on a quantum space-time. This operator has the right semiclassical limit.

However the final expression is not free of ambiguities, the latter being due to the polymer quantization of the scalar field which in the adopted polarization provides an infrared cut-off whose physical interpretation is still elusive and deserves further investigations. 

The implementation of the effective semiclassical dynamics of gravity and the scalar field in a realistic cosmological context (neglecting the polymer parameter) has been considered in \cite{Cianfrani:2015oha} .

\section{Outlook}
QRLG has opened a new perspective on the cosmological sector of LQG, by giving a description of the quantum Universe as a cuboidal lattice with attached spin quantum numbers. It confirmed the main prediction of LQC, namely the bounce replacing the initial singularity, even though the improved regularization scheme still needs to be consistently derived. It is worth noting how the regularization prescription and the semiclassical analysis of full theory are adopted in QRLG, thus a direct connection can be established between them and the choice of the polymer parameter in LQC. In view of the privileged role of $\bar\mu$ scheme in LQC, this will allow to select certain regularizations and semiclassical states within full LQG as soon as a QRLG model able to derive improved regularization will be realized. 

Further developments will concern the investigation on physical states in QRLG, which will allow to trace a complete analogy with the results of LQC, where a clock-like scalar field is present and the scalar constraint is solved on a quantum level. Moreover, other fundamental matter fields will be added and this analysis is expected to provide for the first time some phenomenological implications out of loop quantization, at least in a cosmological setting. 

Probably the hardest technical development will deal with studying the behavior of perturbations in QRLG. One has to consider next-to-the leading order terms within the perturbative expansion around a homogeneous space and this implies the analysis of the nonlocal terms arising in the Hamiltonian after gauge-fixing (similarly to what happens in the radial gauge \cite{Bodendorfer:2015aca} ). This investigation is intriguing in view of the paramount role of perturbations in the era of precision cosmology and also with respect to the definition of a well-grounded paradigm for perturbations in LQC.

\section*{Acknowledgments}

The work of FC was supported by funds provided by the Polish National Science Center under the agreement DEC12
2011/02/A/ST2/00294.
The work of EA was supported by the grant of Polish Narodowe Centrum Nauki nr 2011/02/A/ST2/00300. 
EA wishes to acknowledge the John Templeton Foundation for the supporting grant \#51876. 


\end{document}